\documentclass[10pt,twocolumn,letterpaper]{article}

\usepackage{cvpr}              % To produce the CAMERA-READY version
\usepackage[dvipsnames]{xcolor}

\definecolor{cvprblue}{rgb}{0.21,0.49,0.74}
\usepackage[pagebackref,breaklinks,colorlinks,citecolor=cvprblue]{hyperref}
\usepackage{multirow}
\usepackage{times}
\usepackage{epsfig}
\usepackage{graphicx}
\usepackage{amsmath}
\usepackage{amssymb}
\usepackage{booktabs}
\usepackage[normalem]{ulem}
\useunder{\uline}{\ul}{}
\usepackage{xcolor}
\usepackage{caption}
\usepackage{float}

\def\paperID{197} % *** Enter the Paper ID here
\def\confName{3DV\xspace}
\def\confYear{2026\xspace}

\title{X-LRM: X-ray Large Reconstruction Model for
Extremely Sparse-View Computed Tomography
Recovery in One Second}

\author{Guofeng Zhang$^{1,*}$, Ruyi Zha$^{2,*}$, Hao He$^3$, Yixun Liang$^3$, Alan Yuille$^1$ Hongdong Li$^2$, Yuanhao Cai$^{1,\dagger}$ \\
$^{1}$ Johns Hopkins University, $^{2}$ Australian National University, $^{3}$ HKUST
}

\begin{document}
\maketitle
\begin{abstract}
Sparse-view 3D CT reconstruction aims to recover volumetric structures from a limited number of 2D X-ray projections. Existing feedforward methods are constrained by the scarcity of large-scale training datasets and the absence of direct and consistent 3D representations.
In this paper, we propose an X-ray Large Reconstruction Model (X-LRM) for extremely sparse-view ($<$10 views) CT reconstruction. X-LRM consists of two key components: X-former and X-triplane. X-former can handle an arbitrary number of input views using an MLP-based image tokenizer and a Transformer-based encoder. The output tokens are then upsampled into our X-triplane representation, which models the 3D radiodensity as an implicit neural field.
To support the training of X-LRM, we introduce Torso-16K, a large-scale dataset comprising over 16K volume-projection pairs of various torso organs.
Extensive experiments demonstrate that X-LRM outperforms the state-of-the-art method by 1.5 dB and achieves 27$\times$ faster speed with better flexibility. Furthermore, the evaluation of lung segmentation tasks also suggests the practical value of our approach. Our code and dataset will be released at \url{https://github.com/Richard-Guofeng-Zhang/X-LRM}.
\end{abstract}    
\let\thefootnote\relax\footnotetext{$*$ = Equal Contribution. $\dagger$ = Corresponding Author}
\vspace{-5mm}
\section{Introduction}
\label{sec:intro}

Computed Tomography (CT) uses X-rays with penetrating power to reveal internal structures non-invasively. It is widely used in medical imaging for disease diagnosis, treatment planning, and surgical navigation~\cite{x_ray_1,x_ray_2,x_ray_3,x_ray_4}. In particular, CT reconstruction aims to recover the 3D radiodensity of the scanned object given 2D X-ray projections.

Traditional methods~\cite{fdk,analytical_2,asd_pocs,sart} usually require hundreds of X-ray projections to yield good reconstruction quality, which exposes significant radiation to patients. Recently, some self-supervised algorithms based on neural radiance field (NeRF)~\cite{sax_nerf,zha2022naf} or 3D Gaussian splatting (3DGS)~\cite{cai2024radiative,zha2024r} have been designed to reconstruct CT with $\sim$ 50 projections. Yet, these methods usually require a long time ($\sim$15 minutes) for each reconstruction with still relatively high radiation exposure. In this work, we study the extremely sparse-view ($<$10 views) CT reconstruction in a feedforward manner to inference \textbf{in one second}.

\begin{figure}[t]
    \centering
    \includegraphics[width=1\linewidth]{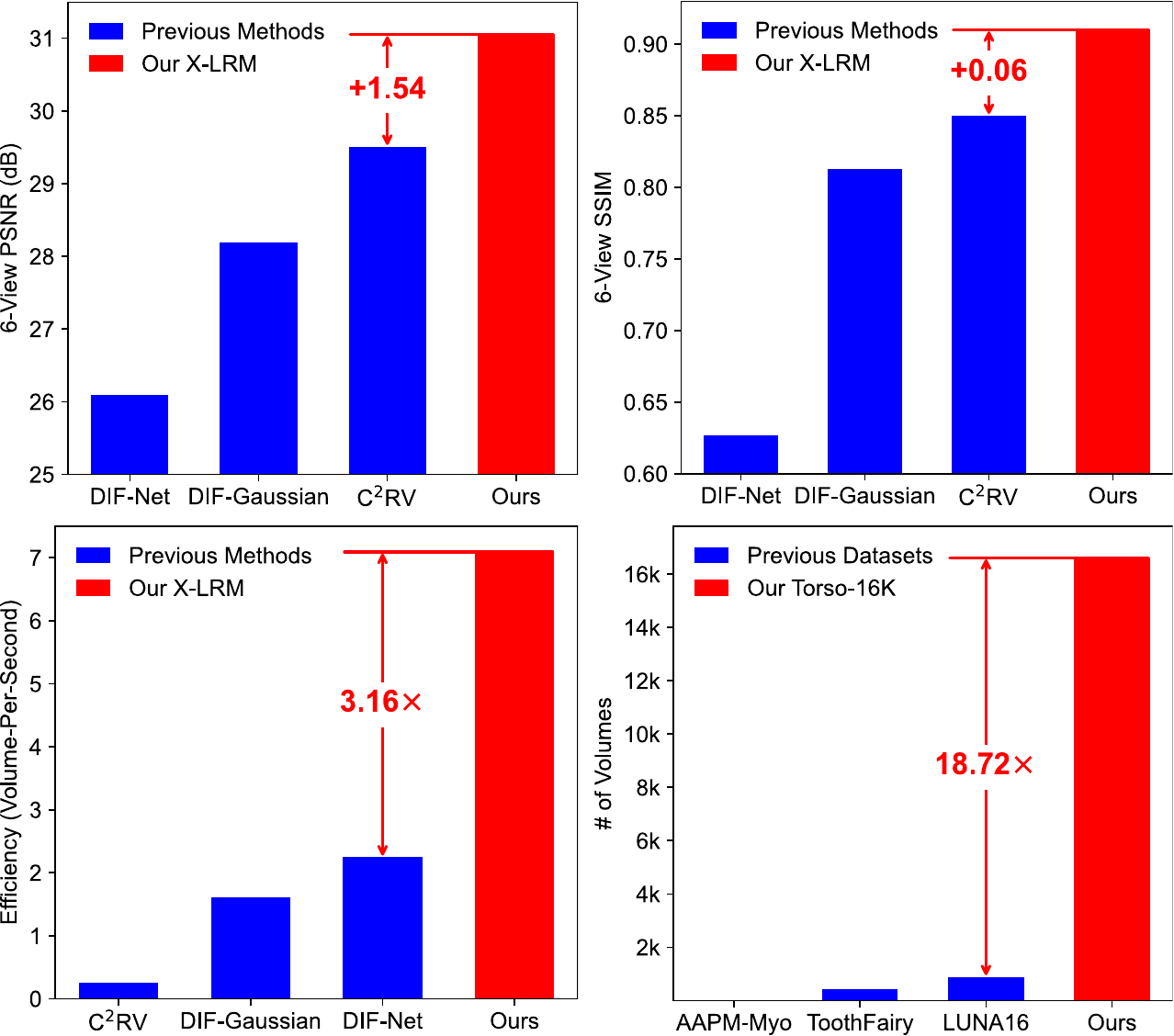}\vspace{-1mm}
    \caption{Our X-LRM outperforms previous 3D feedforward methods in quality and efficiency, including DIF-Net~\cite{difnet}, DIF-Gaussian~\cite{dif_gs}, and C$^2$-RV~\cite{c2rv}. Our collected CT dataset, Torso-16K, is over 18$\times$ larger than previous benchmarks: LUNA16~\cite{setio2017validation}, ToothFairy~\cite{cipriano2022deep}, and AAPM-Myo~\cite{mccollough2016tu}.
    \vspace{-5mm}
}
    \label{fig:teaser}
\end{figure}

Some recent works~\cite{fbpconvnet,c2rv,dif_gs,difnet,freeseed} also try to explore this task. However, existing feedforward methods suffer from the following issues. \textbf{(i)} They rely on single-organ datasets containing fewer than 1,000 cases~\cite{setio2017validation,cipriano2022deep,mccollough2016tu}, which severely lack the diversity and scale required to develop robust and generalizable models.
\textbf{(ii)} The number of the input projections of existing methods is fixed and cannot be adjusted, which lacks flexibility and limits the application in practice. \textbf{(iii)} Previous feedforward methods lack an explicit 3D representation, which limits their ability to model complex spatial structures and hampers performance in sparse-view or large-scale 3D reconstruction settings.

To cope with these problems, we design an X-ray Large Reconstruction Model (X-LRM), for extremely sparse-view ($<$ 10 views) CT recovery. X-LRM consists of two parts: X-former and X-triplane. Firstly, X-former uses a multi-layer perception (MLP) based tokenizer to split an arbitrary number of input images into patch tokens. Then X-former adopts a pure Transformer~\cite{transformer} encoder to compute the self-attention among these patch tokens. The output tokens of the X-former are then upsampled and reshaped into our X-triplane representation. The point feature of the X-triplane is fed into an MLP to learn an implicit neural field of the 3D volume radiodensity. To explore the potential of large-scale training, we collect a 3D CT reconstruction dataset, Torso-16K, containing $\sim$16K volume-projection data pairs. With the proposed techniques and collected dataset, our X-LRM can significantly benefit from large-scale training to boost the reconstruction quality and flexibly handle different numbers of input X-ray projections.

In a nutshell, our contributions can be summarized as:

\begin{itemize}
        \item We propose X-LRM, a novel feedforward framework for sparse-view CT reconstruction.
        \item We design a Transformer-based encoder, X-former, to flexibly encode an arbitrary number of input X-ray projections. Besides, we present a new 3D representation, X-triplane, which directly and consistently models the radiodensity in X-ray imaging.
        \item We collect a large-scale dataset, Torso-16K, containing over 16K samples of 2D X-ray projections and 3D CT volumes. To the best of our knowledge, our Torso-16K is the largest CT reconstruction benchmark and is over 18$\times$ larger than the existing largest dataset.
	\item X-LRM drastically outperforms the SOTA by 1.5 dB PSNR and is $27\times$ faster in inference.
\end{itemize}

\section{Related Works}
\subsection{Sparse-View CT Reconstruction}

We adopt a cone-beam CT (CBCT) setup that acquires multi-view 2D X-ray projections for volumetric reconstruction.
Existing sparse-view CT reconstruction approaches can be categorized into optimization-based and prediction-based methods. 
Optimization-based methods iteratively refine the 3D volume to align with the measured projections. Traditional methods~\citep{sart,sauer1993local,sidky2008image} formulate reconstruction as a \emph{maximum a posteriori} problem, while learning-based methods leverage neural representations~\citep{zha2022naf,shen2022nerp,cai2024structure,zha2024r,cai2024radiative} and diffusion models~\citep{chung2023solving,chung2023decomposed,lee2023improving}. Despite their effectiveness, these methods typically require minutes to hours to process a single case, making them impractical for real-time clinical applications.
Prediction-based methods, in contrast, utilize neural networks to learn semantic priors from external datasets. Given a test case, they employ pre-trained models for projection extrapolation~\citep{anirudh2018lose,ghani2018deep}, slice denoising~\citep{wang2022dudotrans,jin2017deep,freeseed}, or volume regression~\citep{difnet,dif_gs,c2rv}. While these methods enable rapid inference, they are constrained by the limited capacity of CNN-based models and the scarcity of large-scale training datasets. We cope with these problems by designing X-LRM and collecting Torso-16K.
% Our approach belongs to the prediction-based category but overcomes these limitations through key innovations including a tri-plane-based representation, a mixed training strategy, and a large-scale curated dataset.

\subsection{Feedforward 3D Reconstruction}
Unlike optimization-based methods NeRF~\citep{mildenhall2021nerf} or 3D gaussian splatting~\citep{kerbl20233d}, which take time-consuming optimization phase for shape recovery, feedforward 3D reconstruction aims to learn diverse geometry types (\emph{e.g.}, mesh~\citep{wang2018pixel2mesh,wu2020pq}, implicit fields~\citep{mescheder2019occupancy,xu2019disn} \emph{etc.}) from input images in a forward manner with neural network architectures. Boosting from large-scale 3D datasets like Objaverse-XL~\citep{deitke2023objaverse,deitke2023objaversexl} and the scalability of Transformer architectures~\citep {vaswani2017attention}, Large Reconstruction Model~\citep{hong2023lrm} and its subsequent variants~\citep{li2023instant3d,tochilkin2024triposr,wang2023pf,wei2024meshlrm,zhang2024gs,he2024lucidfusion,cai2024baking} has greatly promoted reconstruction ability and efficiency of current fields. However, due to the data scarcity of CT reconstruction and 3D model, current feedforward CT reconstruction methods often suffer from poor reconstruction quality and generalization. Our goal is to fill these research gaps.
\section{Method}
\begin{figure*}
    \centering
    \includegraphics[width=\textwidth,trim=2 2 2 2,clip]{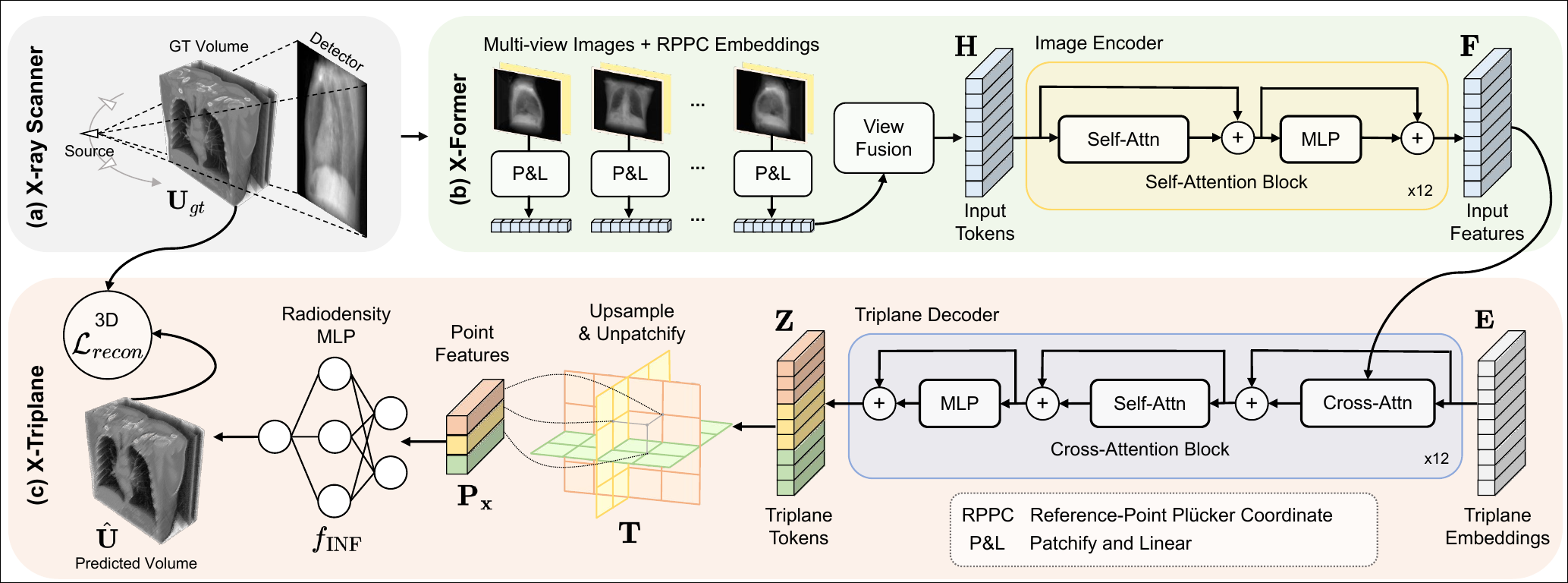}
    \caption{The overall architecture of X-LRM:
       (a) We collect Torso-16K, the largest CT reconstruction dataset (\cref{exp:setup}). 
        (b) Our X-Former features an image tokenizer and encoder, designed to process a variable number of input views (\cref{method:x_former}). 
        (c) Our X-Triplane includes a triplane decoder followed by our implicit neural field, directly predicting the 3D CT volume $\hat{\mathbf{U}}$ (\cref{method:x_triplane}).
        }
    \label{fig:pipeline}
\end{figure*}

The pipeline of our method is shown in \cref{fig:pipeline}. Our X-LRM consists of two parts: X-former and X-triplane, corresponding to \cref{fig:pipeline} (b) and \cref{fig:pipeline} (c). X-former begins with an MLP-based image tokenizer. Then a Transformer-based encoder processes an arbitrary number of multi-view image tokens with view-associated ray information into patch-based features. These features are then mapped into triplane tokens through cross-attention in a triplane decoder. We upsample and unpatchify these tokens to our X-triplane representation. Finally, we adopt an MLP to learn an implicit mapping from the 3D point features on the triplane to the corresponding volume radiodensity.
\subsection{X-former}
\label{method:x_former} 

As aforementioned, existing feedforward methods struggle with large-scale training and varying numbers of projections, resulting in degraded performance, limited scalability, and reduced flexibility. To address these challenges, we propose X-former, an architecture composed of an MLP-based image tokenizer and a Transformer-based image encoder tailored for variable-view processing.

\noindent\textbf{Image Tokenizer.} As shown in \cref{fig:pipeline} (b), the input to the tokenizer is multi-view X-ray projections $\mathbf{I}_i \in \mathbb{R}^{H \times W \times 1}$ concatenated with the corresponding viewpoint camera conditions $\mathbf{C}_i \in \mathbb{R}^{H \times W \times 6}$. We denote the input at the $i$-th view  as $\mathbf{X}_i=[\mathbf{I}_i, \mathbf{C}_i] \in \mathbb{R}^{H \times W \times 7}$. During training, X-former can take varying numbers of views, denoted as $\mathcal{V} = \{V_1, V_2, \dots, V_m\}$, where $m$ is the count of varying numbers of input views.  For a specific $V_i$, the input is denoted as $\mathbf{X}=[\mathbf{X}_1, \mathbf{X}_2, \dots, \mathbf{X}_{V_i}] \in\mathbb{R}^{V_i \times H \times W \times 7}$, where $V_i \in \mathbf{V}$ can change dynamically during training. 

We adopt the Reference-Point Pl"ucker Coordinate (RPPC)~\citep{cai2024baking} as our camera condition, as it encodes more ray position and relative depth information than standard Pl"ucker coordinates. Thus, we have $\mathbf{C}_i = \left( \mathbf{o}_i - (\mathbf{o}_i \cdot \mathbf{d}_i) \mathbf{d}_i, \mathbf{d}_i \right)$, which better captures spatial relationships. Here, $\mathbf{o}_i$ and $\mathbf{d}_i$ denote the origins and directions of pixel-aligned rays at the $i$-th view.

%\paragraph{Tokenizer and Encoder.} As previously mentioned, existing feedforward methods based on CNN architectures struggle to accommodate varying numbers of projections within a single model due to their rigid computational structure, which inherently requires a fixed input size. This significantly hinders their flexibility and scalability in real-world applications, where the number of available projections may vary across different use cases and datasets.

Subsequently, the tokenizer partitions each input view into non-overlapping patches and projects each patch into a latent space of dimension $d_E$ via an MLP layer. Then we fuse patchified tokens of different views by concatenating them to derive the initial patch-wise tokens $\mathbf{H} \in \mathbb{R}^{n \times d_E}$.

\noindent\textbf{Image Encoder.} The feature tokens ${\mathbf{H}}$ are then encoded by a Transformer-based encoder to produce input feature tokens: $\mathbf{F} \in \mathbb{R}^{n \times d_E}$, where $d_E$ is the hidden dimension of our image encoder. The image encoder consists of $N_{e}$ self-attention blocks~\citep{vaswani2017attention}, and each block comprises a multi-head self-attention layer and an MLP layer. We add layer normalization~\citep{ba2016layer} before both layers. For the $j$-th self-attention block, we first split input $\mathbf{H}^j_{in}$ into $k_E$ heads as 
\begin{equation}
    \mathbf{H}^j_{in} = [\mathbf{H}^j_1, \mathbf{H}^j_2, \dots, \mathbf{H}^j_{k_E}].
\end{equation}
Then for the $i$-th head, we project input $\mathbf{H}^j_i$ into $\mathbf{Q}^j_i \in \mathbb{R}^{n \times d_{ke}}$, $\mathbf{K}_i^j \in \mathbb{R}^{n \times d_{ke}}$, and $\mathbf{V}_i^j \in \mathbb{R}^{n \times d_{ke}}$ as
\begin{equation}
\mathbf{Q}_i^j = \mathbf{H}_i^j \mathbf{W}_{\mathbf{Q}_i^j}, \;
\mathbf{K}_i^j = \mathbf{H}_i^j \mathbf{W}_{\mathbf{K}_i^j}, \;
\mathbf{V}_i^j = \mathbf{H}_i^j \mathbf{W}_{\mathbf{V}_i^j},
\end{equation}
where $\mathbf{W}_{\mathbf{Q}_i^j}$, $\mathbf{W}_{\mathbf{K}_i^j}$, $\mathbf{W}_{\mathbf{V}_i^j} \in \mathbb{R}^{d_E \times d_{ke}}$ are learnable parameters of the $fc$ layers and $d_{ke}=d_E / k_E$. Then the output of $i$-th head of the $j$-th self-attention layer $\mathbf{A}_i^j$ is computed as

\begin{equation}
    {\mathbf{A}_i^j} = \text{softmax} \left( \frac{\mathbf{Q}_i^j (\mathbf{K}_i^j)^\top}{\sqrt{d_{ke}}} \right) \mathbf{V}_i^j + \mathbf{H}_i^j.
\end{equation}
Then $k_E$ heads are concatenated to pass through a $fc$ layer to derive the output of self-attention as
\begin{equation}
    {\mathbf{H}^j_{mid}} = [\mathbf{A}_1^j, \; \mathbf{A}_2^j \; \dots \; \mathbf{A}_{k_E}^j ]\; \mathbf{W}^j_s.
\end{equation}
where $\mathbf{W}^j_s \in \mathbb{R}^{d_E \times d_E}$ is the learnable parameter. Then we forward $\mathbf{H}^j_{mid}$ to the MLP layer:
\begin{equation}
    \mathbf{H}^{j}_{out} = \sigma (\mathbf{H}^j_{mid} \mathbf{W}_1 + \mathbf{b}_1) \mathbf{W}_2 + \mathbf{b}_2 + \mathbf{H}^j_{mid},
\end{equation}
where $\sigma$ is the activation function, and $\mathbf{W}_1, \mathbf{W}_2, \mathbf{b}_1, \mathbf{b}_2$ are learnable parameters. The output of the last layer of the image encoder is $\mathbf{F} = \mathbf{H}^{N_e}_{out} \in \mathbb{R}^{n \times d_E}$. This process is illustrated in \cref{fig:pipeline} (b)

Our X-former leverages the inherent flexibility of the transformer architecture, which can naturally process input tokens of different lengths. This allows our model to seamlessly train with different numbers of input views within a single training session, boosting the reconstruction performance and resulting in a unified framework capable of handling diverse multi-view configurations. 

\begin{figure*}[t]
    \centering
    \begin{minipage}{0.45\textwidth}
        \centering
        \label{tab:dataset_stats}
        \renewcommand{\arraystretch}{1.2}
        \resizebox{1\textwidth}{!}{\hspace{3mm}
        \begin{tabular}{@{}lll@{}}
            \toprule[0.15em]
            Dataset & Body Parts & \# of Volumes \\ \midrule
            AbdomenAtlas v1.0 & Abdomen, Chest, Pelvis & 5,171 \\
            RSNA2023 & Abdomen, Pelvis & 4,711 \\
            AMOS & Abdomen & 1,851 \\
            PENGWIN & Pelvis & 100 \\
            TCIA & Abdomen & 833 \\
            MELA & Chest & 1,100 \\
            FLARE24 (subset) & Abdomen, Chest & 1,868 \\
            FUMPE & Chest & 35 \\
            LNDb & Chest & 294 \\ 
            RibFrac & Abdomen, Chest & 660 \\ \midrule
            Torso-16K (Ours) & Abdomen, Chest, Pelvis & 16,623 \\ \bottomrule[0.15em]
        \end{tabular}%
        }
        \captionof{table}{The statistics of our collected Torso-16K benchmark. It integrates ten public datasets covering major anatomical regions in different clinical applications.}
    \end{minipage}
    \hfill
    \begin{minipage}{0.53\textwidth}
        \centering
        % \hspace{-1mm}
        \includegraphics[width=\linewidth]{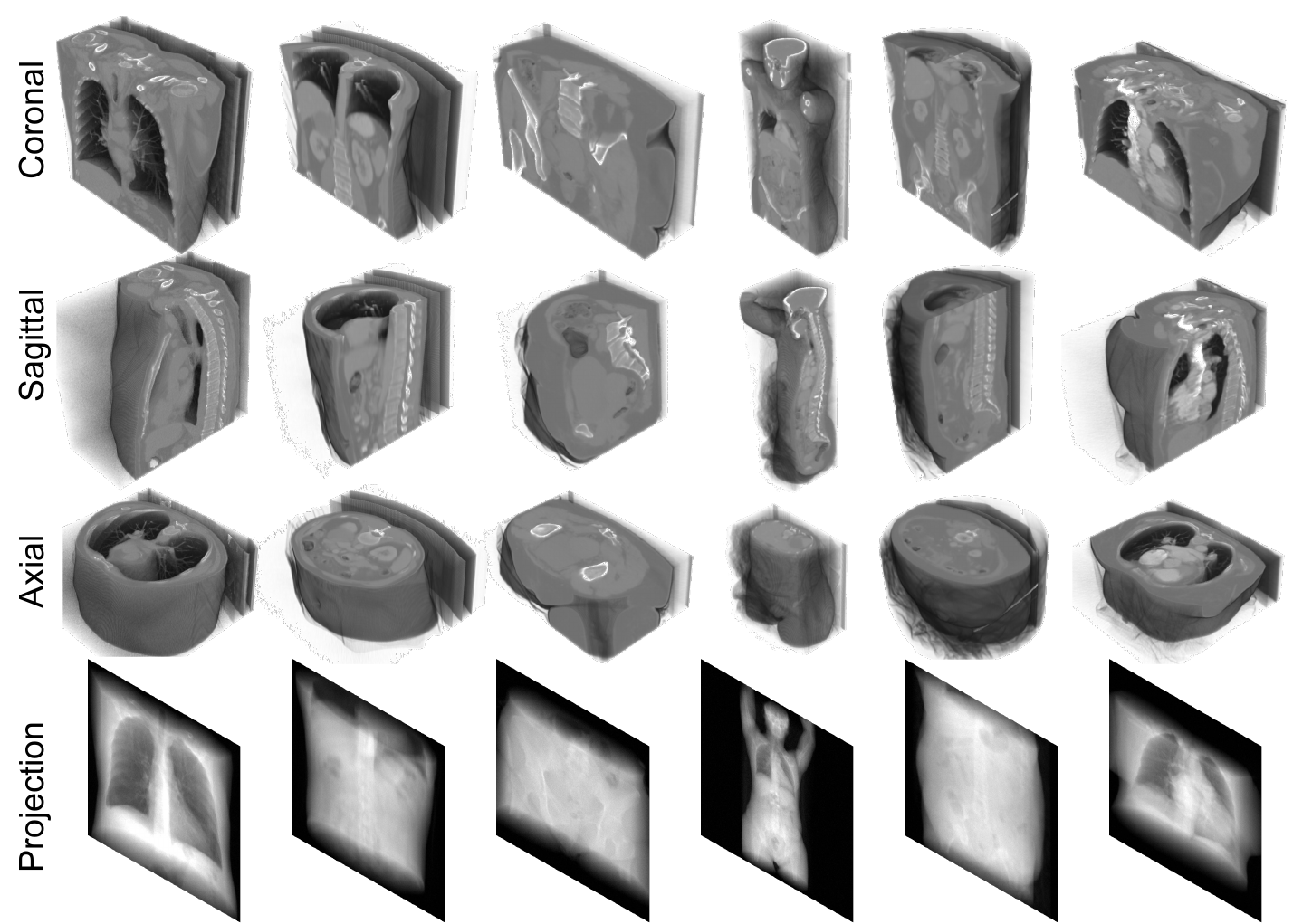}
        \vspace{-6mm}
        \caption{Example CT volumes and corresponding X-ray projections in our Torso-16K dataset.}
        \label{fig:dataset_demo}
    \end{minipage}
\end{figure*}

\subsection{X-triplane}
\label{method:x_triplane}
To lift the features from 2D projection into 3D space, we design a Transformer-based decoder to map the 2D patch-wise features $\mathbf{F}$ into 3D triplane tokens $\mathbf{Z} \in \mathbb{R}^{(3 \times 32 \times 32) \times d_D}$, where $d_D$ is the hidden dimension of the triplane decoder. $\mathbf{Z}$ is later upsampled and reshaped into our X-triplane representations, which encode 3D information. Then we adopt an MLP to learn an implicit mapping from the 3D point feature on the triplane representation to the corresponding radiodensity.

\noindent\textbf{Triplane Decoder.} As shown in \cref{fig:pipeline} (c), the input of the triplane decoder includes $\mathbf{F}$ and a learnable triplane embeddings $\mathbf{E} \in \mathbb{R}^{(3 \times 32 \times 32) \times d_D}$. Our triplane decoder has $N_{d}$ cross-attention blocks. Each cross-attention block comprises a cross-attention layer, a self-attention layer, and an MLP layer. To guide the reconstruction of the triplane tokens and lift the feature into 3D space, we adopt the cross-attention to extract 2D projection and camera information by querying the input feature $\mathbf{F}$.

Similar to self-attention, for the $j$-th cross-attention block in our triplane decoder, we first split $\mathbf{F}$ and input triplane embeddings $\mathbf{E}^j_{in}$ into $k_D$ heads as

\vspace{-2mm}
\begin{equation}
\begin{aligned}
    \mathbf{F} = [\mathbf{F}_1, \mathbf{F}_2, \dots, \mathbf{F}_{k_D}], \mathbf{E}^j_{in} = [\mathbf{E}^j_1, \mathbf{E}^j_2, \dots, \mathbf{E}^j_{k_D}].
\end{aligned}
\end{equation}

Then for the $i$-th cross-attention head, we project $\mathbf{F}_i$ into \textit{query} $\mathbf{Q}_i^j \in \mathbb{R}^{n \times d_{kd}}$, and project $\mathbf{E}_i^j$ into \textit{key} $\mathbf{K}_i^j \in \mathbb{R}^{(3\times32\times32) \times d_{kd}}$ and \textit{value} $\mathbf{V}_i^j \in \mathbb{R}^{(3\times32\times32) \times d_{kd}}$ by three $fc$ layers, where $d_{kd}=d_D / k_D$. Then the output of $i$-th head of the $j$-th cross-attention layer $\mathbf{B}_i^j$ is computed as

\begin{equation}
    {\mathbf{B}_i^j} = \text{softmax} \left( \frac{\mathbf{Q}_i^j (\mathbf{K}_i^j)^\top}{\sqrt{d_{kd}}} \right) \mathbf{V}_i^j + \mathbf{E}_i^j.
\end{equation}
Subsequently, $k_D$ heads are concatenated to pass through an $fc$ layer for the output:
\begin{equation}
    {\mathbf{E}_{mid}^j} = [\mathbf{B}_1^j, \; \mathbf{B}_2^j \; \dots \; \mathbf{B}_{k_D}^j ]\; \mathbf{W}^j_c,
\end{equation}
Similar to previous self-attention (SA) and MLP in \cref{method:x_former} we have
\begin{equation}
    \mathbf{E}^j_{out} = \text{MLP}\big(\text{SA}(\mathbf{E}^j_{mid}) + \mathbf{E}^j_{mid}\big) + \text{SA}(\mathbf{E}^j_{mid}).
\end{equation}

Finally, the triplane decoder outputs $\mathbf{Z} = \mathbf{E}^{N_d}_{out} \in \mathbb{R}^{(3\times32\times32) \times d_D}$, as illustrated in \cref{fig:pipeline} (c). $\mathbf{Z}$ is further upsampled by a deconvolution layer and unpatchified to our X-triplane representation $\textbf{T}$.

\begin{table*}[]
\centering
\resizebox{\textwidth}{!}{%
\begin{tabular}{@{}ccccccccc@{}}
\toprule[0.15em]
\multirow{2}{*}{Type} & \multirow{2}{*}{~~~~~~~~Method~~~~~~~~} & \multirow{2}{*}{~~~~Time (s)$\downarrow$~~~~} & \multicolumn{2}{c}{~~~~~~~~~~~~~~6-View~~~~~~~~~~~~~~} & \multicolumn{2}{c}{~~~~~~~~~~~~~~8-View~~~~~~~~~~~~~~} & \multicolumn{2}{c}{~~~~~~~~~~~~~~10-View~~~~~~~~~~~~~~}  \\ \cmidrule(lr){4-5} \cmidrule(lr){6-7} \cmidrule(lr){8-9}
 &  & & PSNR$\uparrow$ & SSIM$\uparrow$ & PSNR$\uparrow$ & SSIM$\uparrow$ & PSNR$\uparrow$ & SSIM$\uparrow$ \\ \midrule
\multirow{3}{*}{Traditional} & FDK & \textbf{0.008}  & 9.51 & 0.039 & 10.68 & 0.047 & 11.46 & 0.058 \\ 
 & ASD-POCS & 1.385 & 22.17 & 0.573 & 23.40 & 0.612 & 24.62 & 0.667  \\ 
  & SART & 1.400 & 22.61 & 0.537 & 23.56 & 0.548 & 24.57 & 0.585 \\ \midrule
\multirow{2}{*}{2D Feedforward} & FBPConvNet  & {\ul 0.010} & 26.99 & 0.704 & 27.22 & 0.722 & 28.05 & 0.737 \\
 & FreeSeed & 0.163 & 28.93 & 0.841 & {\ul 30.08} & 0.843 & 30.17 & 0.855 \\ \midrule
\multirow{4}{*}{3D Feedforward} & DIF-Net & 0.445 & 26.10 & 0.627 & 26.81 & 0.663 & 27.47 & 0.708 \\
 & DIF-Gaussian & 0.621 & 28.19 & 0.813 & 28.53 & 0.820 & 29.52 & 0.848 \\
 & C$^2$RV & 3.837 & {\ul 29.51} & {\ul 0.850} & 29.83 & {\ul 0.849} & {\ul 30.96} & {\ul 0.871} \\
 & X-LRM (Ours) & 0.141 & \textbf{31.05} & \textbf{0.910} & \textbf{31.24} & \textbf{0.912} & \textbf{31.33} & \textbf{0.915} \\ \bottomrule[0.15em]
\end{tabular}%
}\vspace{-2mm}
\caption{Comparison with traditional and feedforward methods on 750 test cases. X-LRM is 1.5 dB better and 27$\times$ faster than the best baseline. Best result is in \textbf{bold} and second-best is \ul{underlined}.}
\label{tab:750test}
\end{table*}

\begin{figure*}[h]
    \centering
    \hspace{-4mm}
    \includegraphics[width=1.01\textwidth]{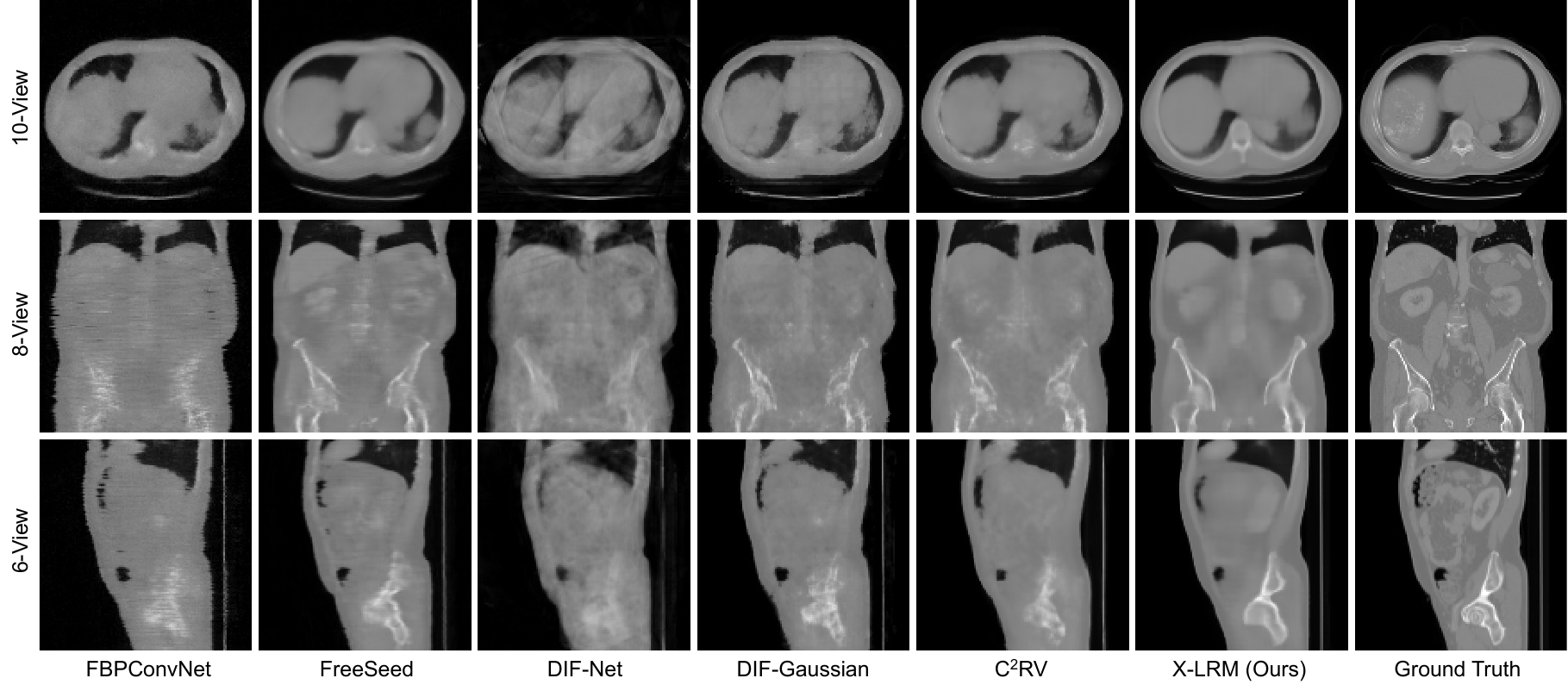}
    \vspace{-2mm}
    \caption{Qualitative results of feedforward methods across multiple anatomical views on the 750-sample test set. From top to bottom: 10-view axial, 8-view coronal, and 6-view sagittal slices. }

    \vspace{-5mm}
    \label{fig:qualitative_feedforward}
\end{figure*}

\noindent\textbf{Triplane Implicit Neural Field.} Our X-triplane $\mathbf{T}$ is composed of three orthogonal feature planes: $\mathbf{T}{xy}$, $\mathbf{T}{yz}$, and $\mathbf{T}{xz} \in \mathbb{R}^{(64 \times 64) \times d_T}$, where $64 \times 64$ refers to the spatial resolution of each plane and $d_T$ is the dimension of the point feature $\mathbf{P}{\boldsymbol{x}} \in \mathbb{R}^{3 \times d_T}$. Then we build an implicit neural field mapping from the position and feature of a 3D point to its radiodensity.

For a given 3D point $\boldsymbol{x} = (x,y,z) \in [-1,1]^3$ within the unit bounding box (where each coordinate is normalized), we obtain its feature embeddings by projecting it onto three orthogonal plane features $\mathbf{T}_{xy}, \mathbf{T}_{yz},$ and $\mathbf{T}_{xz}$ at $\mathbf{p}_{xy} = (x, y), \mathbf{p}_{yz} = (y, z),$ and $ \mathbf{p}_{xz} = (x, z)$. We then apply bilinear interpolation to extract features from each plane. Take the $xy$-plane $\mathbf{T}_{xy}$ and a point $\mathbf{p}_{xy} = (x, y)$ for instance, the interpolated feature value is computed as
\begin{equation}
    \begin{aligned}
        \mathbf{T}_{xy}(\mathbf{p}_{xy}) &= 
        (1 - \alpha) \beta \mathbf{T}(x_0, y_1) + \alpha (1 - \beta) \mathbf{T}_{xy}(x_1, y_0) \\
        & + (1 - \alpha)(1 - \beta) \mathbf{T}_{xy}(x_0, y_0) + \alpha \beta \mathbf{T}_{xy}(x_1, y_1),
    \end{aligned}
\end{equation}
where $x_0, x_1$ and $y_0, y_1$ are the neighboring points, and the interpolation weights are $\alpha = x - x_0$, $\beta = y - y_0$. Applying to all three triplanes, we obtain the feature at the point $\boldsymbol{x}$ as
\begin{equation}
    \mathbf{P}_{\boldsymbol{x}} = \left( \mathbf{T}_{xy}(\mathbf{p}_{xy}), \mathbf{T}_{yz}(\mathbf{p}_{yz}), \mathbf{T}_{xz}(\mathbf{p}_{xz}) \right).
\end{equation}

As the radiodensity is isotropic and only related to the point property,  we adopt an MLP to learn the mapping $f_{\text{INF}}$ from the point feature $\mathbf{P}_{\boldsymbol{x}}$ to the radiodensity $\rho_{\boldsymbol{x}}$ as %we feed the point feature $\mathbf{P}_{\boldsymbol{x}}$ to an MLP to learn the mapping $f_{\text{INF}}$ to the radiodensity $\rho_{\boldsymbol{x}}$ as

\begin{equation}
    f_{\text{INF}}:\left( \mathbf{T}_{xy}(\mathbf{p}_{xy}), \mathbf{T}_{yz}(\mathbf{p}_{yz}), \mathbf{T}_{xz}(\mathbf{p}_{xz}) \right) \rightarrow \rho_{\boldsymbol{x}}.
\end{equation}

\subsection{Training Objective}
\label{method:training_objective}
Existing RGB 3D reconstruction methods mainly adopt 2D rendering training loss to achieve good image recovery quality. However, this supervision involves volume rendering that needs to sample many 3D points to compute for each ray, taking a long time and increasing memory cost. Besides, in X-ray imaging, the 3D CT reconstruction is more concerned than the 2D X-ray rendering. Thus, we adopt the more precise 3D reconstruction loss with varying numbers of input views as
\vspace{-1mm}
\begin{equation}
    \mathcal{L}_{recon} = \frac{1}{m}\sum_{V_i \in \mathcal{V}} \big|\big|~\hat{\mathbf{U}}_{V_i} - \mathbf{U}_{gt} \ \big|\big|^2,
\end{equation}
\vspace{-1mm}
where $\mathcal{V} = \{V_1, V_2, \dots, V_m\}$ represents training settings with different input view numbers $V_i$. $\hat{\mathbf{U}}_{V_i}$ refers to the CT volume reconstructed by X-LRM given $V_i$ views, and $\mathbf{U}_{gt}$ is the ground-truth CT volume. Such 3D supervision enables better anatomical consistency to view sparsity.
\vspace{-2mm}

\begin{table*}[]
\centering
\resizebox{\textwidth}{!}{%
\begin{tabular}{@{}ccccccccc@{}}
\toprule[0.15em]
\multicolumn{1}{l}{\multirow{2}{*}{Type}} & \multirow{2}{*}{~~~~~~~~~~~~Method~~~~~~~~~~~~} & \multirow{2}{*}{~~~~~~~~Time$\downarrow$~~~~~~~~}  & \multicolumn{2}{c}{~~~~~~~~~~~~6-View~~~~~~~~~~~~} & \multicolumn{2}{c}{~~~~~~~~~~~~8-View~~~~~~~~~~~~} & \multicolumn{2}{c}{~~~~~~~~~~~~10-View~~~~~~~~~~~~} \\ \cmidrule(lr){4-5} \cmidrule(lr){6-7} \cmidrule(lr){8-9}
\multicolumn{1}{l}{} &  &  & PSNR$\uparrow$ & SSIM$\uparrow$ & PSNR$\uparrow$ & SSIM$\uparrow$ & PSNR$\uparrow$ & SSIM$\uparrow$ \\ \midrule
\multirow{3}{*}{Self-Supervised} & NAF & 11m & 23.86 & 0.644 & 24.64 & 0.654 & 25.38 & 0.685 \\
 & R$^2$-Gaussian & {\ul 6m} & 20.28 & 0.528 & 20.79 & 0.529 & 22.09 & 0.581  \\
 & SAX-NeRF & 8h & 24.08 & 0.669 & 24.73 & 0.674 & 25.68 & 0.692 \\ \midrule
\multirow{2}{*}{Diffusion Based} 
 & DDS & 12m & 24.42 & 0.529 & 25.64 & 0.570 & 26.64 & 0.607 \\ 
 & DiffusionMBIR & 11h & {\ul 26.61} & {\ul 0.734} & {\ul 28.51} & {\ul 0.803} & {\ul 30.05} & {\ul 0.835} \\ \midrule
3D Feedforward & X-LRM (Ours) & \textbf{0.14s} & \textbf{30.14} & \textbf{0.888} & \textbf{30.10} & \textbf{0.886} & \textbf{30.28} & \textbf{0.889}  \\ \bottomrule[0.15em]
\end{tabular}%
}\vspace{-2mm}
\caption{Comparison with self-supervised and diffusion-based methods on 10 test cases. Our X-LRM achieves 3.53 dB higher PSNR and is 2570$\times$ faster than the best-performing baseline.}
\vspace{-1mm}
\label{tab:10test}
\end{table*}

\begin{figure*}[h]
    \centering
    \includegraphics[width=\textwidth]{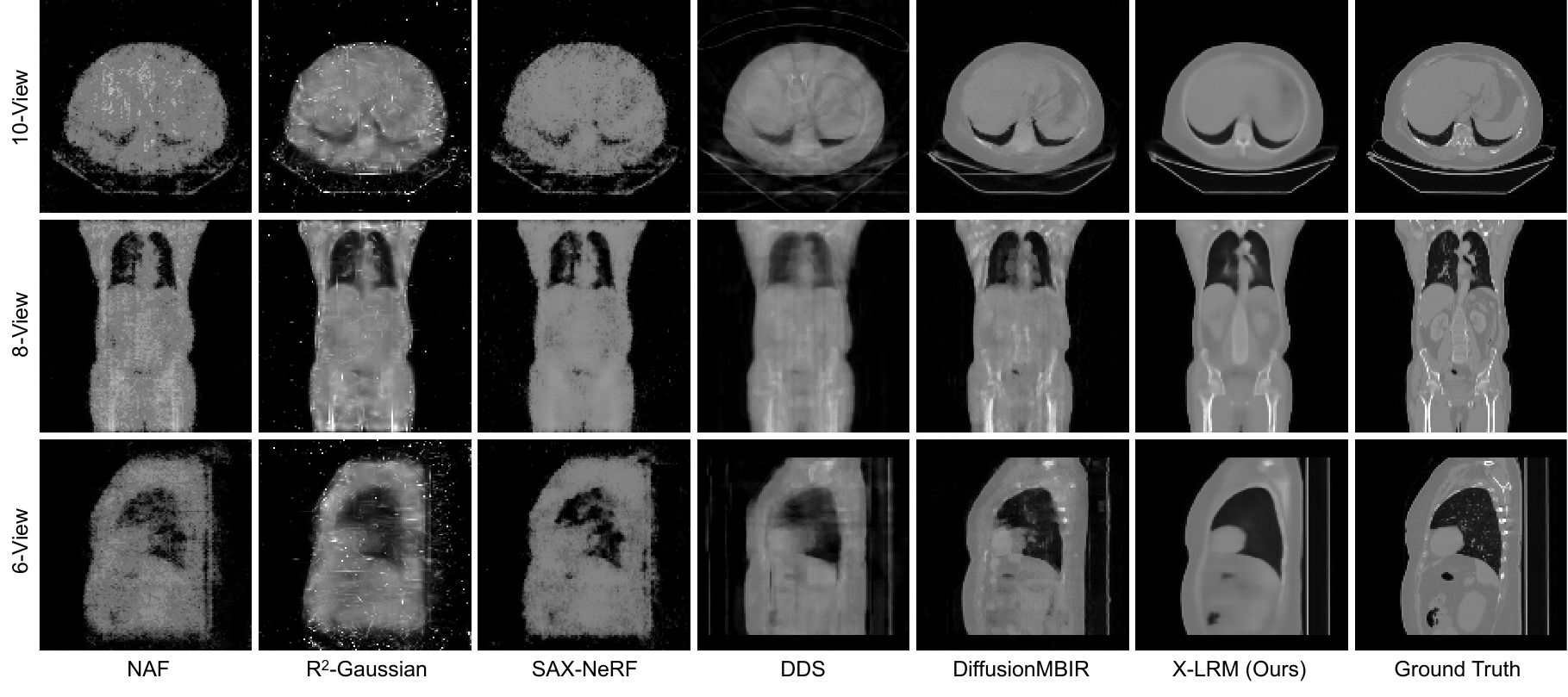}
    \vspace{-5mm}
    \caption{
    Qualitative results of self-supervised and diffusion-based methods on the 10-sample test set. From top to bottom: 10-view axial, 8-view coronal, and 6-view sagittal slices.}
    \vspace{-5mm}
    \label{fig:qualitative_optimization}
\end{figure*}

\section{Experiments}
\vspace{-1mm}
\subsection{Experiment Setup}
\label{exp:setup}
\paragraph{Datasets.}
\vspace{-1mm}
Previous works rely on small datasets~\citep{mccollough2016tu,cipriano2022deep,setio2017validation} (fewer than 1,000 samples), which limits the ability to train generalizable models. To overcome this constraint, we introduce Torso-16K, the largest and most diverse CT reconstruction dataset, comprising 16,623 real-world CT scans from ten public datasets (\cref{tab:dataset_stats}). It covers key anatomical regions in clinical applications, including chest, abdomen, and pelvis. Some examples are shown in~\cref{fig:dataset_demo}. Torso-16K is split into 15,000 / 873 / 750 for training, validation, and testing.

We standardize CT scans by resampling and cropping to a 50$^{3}$ cm$^3$ volume at 128$^{3}$ resolution. Radiodensity values are normalized from the Hounsfield unit range [-1000, 1000] to [0,1], ensuring coverage of primary organs of interest. Since most public datasets only provide CT volumes, we render multi-view X-ray projections via TIGRE toolbox~\citep{biguri2016tigre}. 256$^{2}$ resolution projections span the full range 0$^{\circ}\sim$360$^{\circ}$ with 3.9$^{2}$ mm$^2$ pixel spacing. To enhance realism, we add Gaussian and Poisson noise to simulate Compton scattering and adopt UCT 960+ scanner~\citep{uct960plus}, with 0.6m source-object and 1.118m source-detector distance.

\vspace{-3mm}
\paragraph{Implementation Details.}
We implement X-LRM by PyTorch~\citep{pytorch}. X-LRM is trained with the AdamW optimizer~\citep{loshchilov2017adamw} ($\beta_1 = 0.9$, $\beta_2 = 0.95$). The initial learning rate is set to $4 \times 10^{-4}$ and follows a cosine annealing scheduler~\citep{loshchilov2016sgdr} with a warm-up phase of $3000$ iterations. For the network architecture, we utilize a ViT-B/16 transformer encoder, which processes $256 \times 256$ inputs to $257$ feature tokens at an embedding dimension of $d_E = 384$ with $N_e = 12$ layers. The transformer decoder consists of $N_d = 12$ layers at an output dimension of $d_D = 512$, while the X-triplane has a feature dimension of $d_T = 32$. The MLP used for radiodensity queries has four layers with a hidden dimension of $64$.

During training, our model is designed to learn from a set of possible input view counts, $V = \{6, 8, 10\}$. For each epoch, the same instance is processed 3 times, each with a different number of views selected from $V$. 
% However, within each batch, all samples share the same number of input views. 
Training is conducted on 8 RTX A5000 GPUs at a per-gpu batch size of $6$ for $100$ epochs. For evaluation, we adopt the peak signal-to-noise ratio (PSNR) and the structural similarity index measure (SSIM)~\citep{ssim} as the quantitative metrics. Please note that PSNR is measured directly in 3D space and SSIM is computed as the average of 2D SSIM values.

\begin{figure*}
\begin{minipage}{0.49\textwidth}
\centering
\renewcommand{\arraystretch}{1.1}
\resizebox{\textwidth}{!}{
\begin{tabular}{ccccccc}
\toprule[0.15em]
\multirow{2}{*}{Method} & \multicolumn{2}{c}{Recon.} & \multicolumn{2}{c}{Left Lung} & \multicolumn{2}{c}{Right Lung} \\
\cmidrule(lr){2-3} \cmidrule(lr){4-5} \cmidrule(lr){6-7}
& PSNR & SSIM & DICE  & ASD$\downarrow$ & DICE & ASD$\downarrow$ \\
\midrule
FDK & 9.14 & 0.03 & 0.34 & 43.41 & 0.26 & 45.12 \\
SART & 21.7 & 0.51 & 28.29 & 13.44 & 2.92 & 28.12 \\
ASD-POCS & 21.48 & 0.53 & 25.35 & 15.62 & 2.52 & 31.84 \\
FBPConvNet & 26.02 & 0.68 & {\ul 93.59} & {\ul 0.65} & {\ul 93.58} & {\ul 0.56} \\
FreeSeed & 27.77 & {\ul 0.83} & 91.01 & 1.07 & 90.56 & 0.82 \\
DIF-Net & 24.71 & 0.55 & 84.63 & 1.70 & 84.78 & 1.44 \\
DIF-Gaussian & 26.84 & 0.79 & 92.16 & 0.83 & 91.69 & 0.72 \\
C$^2$RV & {\ul 28.24} & {\ul 0.83} & 91.47 & 0.88 & 90.28 & 0.87 \\
\midrule
X-LRM (Ours) & \textbf{30.59} & \textbf{0.92} & \textbf{95.21} & \textbf{0.49} & \textbf{94.63} & \textbf{0.48} \\
\bottomrule[0.15em]
\end{tabular}
}
\captionof{table}{Traditional and feedforward methods.}
\vspace{-2mm}
\label{tab:lungseg_feedforward}
\end{minipage}
\hfill
\begin{minipage}{0.48\textwidth}
\centering
% Top-right table
\renewcommand{\arraystretch}{1.5}
{\resizebox{\textwidth}{!}{
\begin{tabular}{ccccccc}
\toprule[0.15em]
\multirow{2}{*}{Method} & \multicolumn{2}{c}{Recon.} & \multicolumn{2}{c}{Left Lung} & \multicolumn{2}{c}{Right Lung} \\
\cmidrule(lr){2-3} \cmidrule(lr){4-5} \cmidrule(lr){6-7}
& PSNR & SSIM & DICE  & ASD$\downarrow$ & DICE & ASD$\downarrow$ \\
\midrule
NAF & 21.91 & 0.57 & 48.54 & 19.14 & 50.08 & 9.49 \\
R$^2$-Gaussian & 18.58 & 0.45 & 29.62 & 26.12 & 34.76 & 12.86 \\
SAX-NeRF & 21.83 & 0.59 & 39.34 & 27.55 & 19.87 & 20.86 \\
DiffusionMBIR & {\ul 25.31} & {\ul 0.72} & {\ul 93.10} & {\ul 0.74} & {\ul 93.25} & {\ul 0.67} \\
DDS & 23.47 & 0.53 & 71.04 & 2.61 & 69.58 & 2.60 \\
\midrule
X-LRM (Ours) & \textbf{27.63} & \textbf{0.85} & \textbf{95.60} & \textbf{0.51} & \textbf{95.48} & \textbf{0.48} \\
\bottomrule[0.15em]
\end{tabular}
}}
\captionof{table}{Self-supervised and diffusion methods.}
\label{tab:lungseg_opt}
\end{minipage}
\end{figure*}

\begin{figure*}[ht]
    \centering
    \includegraphics[width=\textwidth]{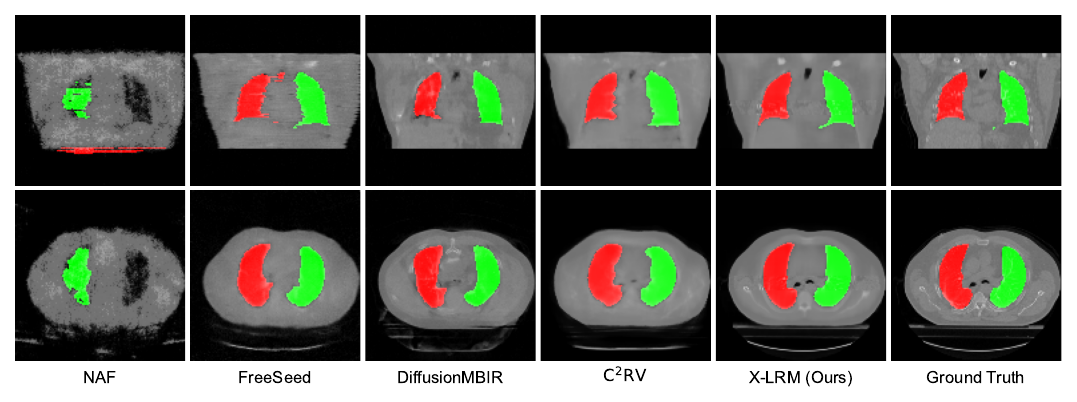}
    \vspace{-8mm}
    \caption{Visual comparison of lung segmentation on 6-view reconstructed CT slices with the recent best self-supervised method NAF~\cite{zha2022naf}, 2D feedforward method FreeSeed~\cite{freeseed}, diffusion-based method DiffusionMBIR~\cite{chung2023solving}, and 3D feedforward method C$^2$RV~\cite{c2rv}.}
    \label{fig:lung_seg_vis}
    \vspace{-2mm}
\end{figure*}

\subsection{Comparison with State-of-the-Art Methods}
We evaluate our X-LRM model against baseline methods under different numbers of projection views (\emph{i.e.} 6, 8, 10) using the following two different settings:

\begin{itemize}
\item \textbf{Traditional and feedforward methods}: Traditional methods are directly tested on the 750-sample test set. The 2D and 3D feedforward methods are first trained on the train set and then tested on the 750-sample test set.
\item \textbf{Self-supervised and diffusion-based methods}: We use a subset of 10 samples selected from the 750-sample test set, ensuring all 10 sub-datasets in \cref{tab:dataset_stats} are covered. We test on this small dataset due to the long inference time of these methods.
\end{itemize}

\noindent\textbf{Quantitative Results.} Firstly, we compare X-LRM with three traditional methods (FDK~\citep{fdk}, SART~\citep{sart}, and ASD-POCS~\citep{asd_pocs}) and five feedforward methods (FBPConvNet~\citep{fbpconvnet}, FreeSeed~\citep{freeseed}, DIF-Net~\citep{dif_gs}, DIF-Gaussian~\citep{dif_gs}, and C$^2$RV~\citep{c2rv}). The results are reported in Tab.~\ref{tab:750test}. \textbf{(i)} When reconstructing CT volumes from 6, 8, and 10 X-ray projection views, our X-LRM surpasses the SOTA 2D feedforward method, FreeSeed, by 2.12, 1.16, and 1.16 dB in PSNR. Compared to the SOTA 3D feedforward method, C$^2$RV, X-LRM improves the performance by 1.54, 1.41, and 0.37 dB in PSNR, while enjoying over 27$\times$ faster inference speed. \textbf{(ii)} Unlike previous feedforward methods, X-LRM enjoys better flexibility as it can efficiently reconstruct CT volume with different numbers of input views without training separate models.

Secondly, we compare with three self-supervised methods (NAF~\citep{zha2022naf}, R$^2$-Gaussian~\citep{r2_gaussian}, and SAX-NeRF~\citep{sax_nerf}) and two diffusion-based methods (DiffusionMBIR~\citep{chung2023solving} and DDS~\citep{chung2023decomposed}). The quantitative results are listed in Tab.~\ref{tab:10test}. Our X-LRM achieves the best performance and fastest inference speed. Compared with the second-best method, DiffusionMBIR, our X-LRM is 3.53 dB higher in PSNR. Compared with the second-fastest method, R$^2$-Gaussian, our method is over 2570$\times$ faster in inference.

\vspace{-4mm}

\paragraph{Qualitative Results.}
The qualitative results are depicted in  \cref{fig:qualitative_feedforward} (compared with feedforward methods) and \cref{fig:qualitative_optimization} (compared with self-supervised and diffusion-based methods). As observed from the reconstructed slices, all baseline methods struggle with generating high-quality reconstructions, particularly in sparser-view scenarios. Both feedforward and optimization-based approaches exhibit noticeable blurriness and lack of fine details, leading to incomplete anatomical structures and texture inconsistencies. Structural elements, such as lung regions and organ boundaries, appear unclear, often blending into surrounding areas due to the loss of high-frequency details.

\begin{table}[h]
    \centering
    \resizebox{1\columnwidth}{!}{
    \begin{tabular}{cccc}
        \toprule[0.15em]
        Method &~~Base Model~~ &~~+ X-Triplane~~ &~~+ X-former~~ \\
        \midrule
        PSNR  &13.09  &28.76  &\bf 31.33 \\
        SSIM  &0.42  &0.84  &\bf 0.92 \\
        \bottomrule[0.15em]
    \end{tabular}
}\vspace{0mm}
\caption{Break-down ablation towards higher performance by adding the components of X-LRM. The ablation study is conducted under the 10-view CT reconstruction setting. }
\label{tab:breakdown}
\vspace{-2mm}
\end{table}

\begin{table}[]
\centering
\resizebox{\columnwidth}{!}{%
\begin{tabular}{ccccc}
\toprule[0.15em]
\multicolumn{3}{c}{Noisy parameters} & \multirow{2}{*}{PSNR} & \multirow{2}{*}{SSIM($10^{-2}$)} \\
\cmidrule(lr){1-3}
Angles  & DSO  & DSD  &  &  \\
\midrule
- & - & - & 31.05 \textcolor{gray}{(-0.00)} & 91.04 \textcolor{gray}{(-0.00)} \\
\midrule
$\pm 0.5^\circ$   & \multirow{2}{*}{-} & \multirow{2}{*}{-} & 30.93 \textcolor{gray}{(-0.12)} & 90.95 \textcolor{gray}{(-0.09)} \\
$\pm 1^\circ$     &        &      & 30.62 \textcolor{gray}{(-0.43)} & 90.71 \textcolor{gray}{(-0.33)} \\
\midrule
\multirow{2}{*}{-} & $\pm 2mm$   & \multirow{2}{*}{-} & 30.85 \textcolor{gray}{(-0.20)} & 90.89 \textcolor{gray}{(-0.15)} \\
 & $\pm 3mm$  &       & 30.67 \textcolor{gray}{(-0.38)} & 90.73 \textcolor{gray}{(-0.31)} \\
\midrule
\multirow{2}{*}{-} & \multirow{2}{*}{-} & $\pm 2mm$  & 30.99 \textcolor{gray}{(-0.06)} & 90.99 \textcolor{gray}{(-0.05)} \\
   &       & $\pm 3mm$  & 30.93 \textcolor{gray}{(-0.12)} & 90.95 \textcolor{gray}{(-0.09)} \\
\bottomrule[0.15em]
\end{tabular}%
}\vspace{0mm}
\caption{Ablation study of X-LRM's robustness to noisy X-ray scanner parameters under a 6-view CT reconstruction setting.}
\label{tab:noise_ablate}
\vspace{-5mm}
\end{table}

In contrast, our X-LRM yields visually sharper reconstructions with well-defined textures and more coherent anatomical structures. Across different view settings, it preserves fine-grained details while maintaining spatial smoothness. Our method consistently reconstructs realistic features with minimal artifacts, demonstrating high-quality performance in sparse-view CT reconstruction.

\vspace{-2.5mm}
\paragraph{Application in Segmentation.}
We evaluate the reconstructed CT volumes using medical segmentation. We employ the LungMask toolkit~\citep{hofmanninger2020automatic} to segment the left and right lung from CT reconstructions produced by various methods and compare the results against the ground-truth segmentation obtained from the original CT scans. Specifically, we evaluate lung test data from the 750-test set and 10-test set, testing the corresponding baseline methods and reporting reconstruction performance (PSNR and SSIM) alongside lung segmentation accuracy (DICE and ASD) for 6-view reconstructed volumes. As shown in \cref{tab:lungseg_feedforward} and \cref{tab:lungseg_opt}, X-LRM achieves superior reconstruction quality, surpassing C$^2$RV by and DiffusionMBIR by 2.35 and 2.32 dB in PSNR. Additionally, the higher DICE scores and lower ASD values on both the left and right lung indicate that the 3D segmentation on the CT volume reconstructed by X-LRM has a larger overlap and smaller boundary discrepancies with the segmentation mask on the ground-truth CT volume. \cref{fig:lung_seg_vis} shows the visual comparison with four kinds of recent best methods. Both quantitative and qualitative results demonstrate the ability of X-LRM to preserve anatomical structures more accurately and maintain precise shape consistency, surpassing other methods in both reconstruction fidelity and segmentation alignment.

\vspace{-1mm}
\subsection{Ablation Study}
\vspace{-1mm}
Ablation studies evaluate the effectiveness of the proposed modifications compared to the standard LRM, including X-former and X-triplane. Additionally, we assess the robustness of X-LRM under varying noisy scanning parameters, such as viewing angles, DSD, and DSO. The breakdown study is performed under 6,8,10-view setting, and the robustness analysis is conducted under 6-view setting.

\vspace{-4mm}
\paragraph{Break-down Ablation.}
We adopt the Open-LRM~\citep{hong2023lrm} as the base model to study the effect of each component of X-LRM towards higher performance. The results of the 10-view reconstruction are reported in Tab.~\ref{tab:breakdown}. The base model only achieves poor results of 12.33 dB in PSNR on average. After applying our X-triplane and X-former, the model gains by 15.53 and 3.34 dB in PSNR on average. These results validate the effectiveness of our proposed methods.

\vspace{-4mm}
\paragraph{Robustness Analysis.}
We conduct a robustness analysis under varying noisy scanning parameters, including viewing angles, source-to-origin distance (DSO), and source-to-detector distance (DSD). The introduced noise follows a uniform distribution, modeled as $\eta \sim \mathcal{U}(-\epsilon, +\epsilon)$. With this noise, the projection images change but the model processes them as if captured under perfect conditions. Tab.~\ref{tab:noise_ablate} shows that X-LRM remains robust to noises of scanning parameters. Viewing angle shifts of $\pm0.5^\circ$ (PSNR -0.12 dB, SSIM -0.0009) and $\pm1^\circ$ results (PSNR -0.43 dB, SSIM -0.0033) have minimal impact. Noises in DSO and DSD only introduce minor effects, demonstrating the reliability of X-LRM under different real-world possible noises.
\section{Conclusion}
In this paper, we collect the largest dataset, Torso-16K, to enable large-scale training for CT reconstruction. Torso-16K is over 18× larger than the existing largest benchmark. We propose X-LRM, a Transformer-based feedforward framework consisting of X-former and X-triplane. X-former employs a tokenizer and Transformer backbone to flexibly encode an arbitrary number of input views, enabling X-LRM to reconstruct CT volumes without re-training. X-triplane decodes image tokens into a triplane representation and learns a neural implicit function to model 3D radiodensity. Experiments show that X-LRM surpasses the SOTA 3D feedforward method by 1.5 dB while achieving 27× faster speed, with its application in medical segmentation further highlighting its practical value.
\vspace{-2mm}
\section*{Acknowledgement}
This work was supported by the Lustgarten Foundation for Pancreatic Cancer Research, the Patrick J. McGovern Foundation Award, and the National Institutes of Health (NIH) under Award Number
R01EB037669.

{
    \small
    \bibliographystyle{ieeenat_fullname}
    \bibliography{main}
}

\end{document}